\documentstyle[prl,aps,epsf]{revtex}
\draft
\begin{document}
\twocolumn[\hsize\textwidth\columnwidth\hsize\csname @twocolumnfalse\endcsname

\title{Rectification and Flux Reversals for Vortices 
Interacting with Triangular Traps} 
\author{C.J. Olson Reichhardt and C. Reichhardt} 
\address{ 
Center for Nonlinear Studies and Theoretical Division, 
Los Alamos National Laboratory, Los Alamos, New Mexico 87545}

\date{\today}
\maketitle
\begin{abstract}
We simulate vortices in superconductors interacting
with two-dimensional arrays of triangular traps. 
We find that, upon application of an ac drive, 
a net dc flow can occur which shows current reversals with increasing
ac drive amplitude for certain vortex densities,
in agreement with recent experiments and theoretical
predictions. 
We identify the vortex dynamics responsible for the different
rectification regimes.
We also predict the occurrence of a novel transverse rectification effect  
in which a dc flow appears that is transverse to the direction 
of the applied ac drive.   
\end{abstract}
\vspace{-0.1in}
\vspace{-0.3in}

\vskip2pc]
\narrowtext
Vortices interacting with periodic pinning structures have been 
attracting considerable attention 
since they are an ideal system in which to study 
the static and dynamical behaviors
of collectively interacting particles coupled to periodic substrates. 
In these systems, 
a variety of commensurability effects and novel vortex crystals can occur
as a function of the ratio of the number of vortices to the
number of pinning sites \cite{Harada1,Olson2,Martin3}.
The vortex crystals are composed of one or more vortices trapped
at each pin with any remaining vortices located in the 
interstitial regions between the pinning sites.
A variety of dynamical flow phases occur
when a drive is applied, 
including the flow of interstitial vortices 
between the pinning sites \cite{Reichhardt4,Look5} as well as channeling along 
rows of pinning \cite{Reichhardt4}.     
Much of the physics of vortices interacting with periodic pinning is 
also observed 
for repulsively interacting colloids in 
periodic optical trap arrays \cite{Grier6,Brunner7,Bechinger8}.  
In addition to the basic science issues, 
vortices interacting with periodic pinning arrays 
are relevant to possible  
technological applications of superconductors, 
such as critical current enhancement or controlled 
motion of flux for new types of devices. 

Several methods have been proposed for using periodic pinning or 
controlled disorder in superconductors to 
create vortex ratchets, rectifiers, and 
logic devices \cite{Janko9,Janko210,Hastings111,Hastings12,Zhu13}.  
As in general ratchet systems, 
which can be deterministic or stochastic in nature \cite{Review14}, 
a vortex ratchet transforms
an ac input into a dc response.
Vortex ratchets can be created via
asymmetric periodic pinning lines \cite{Janko9,Janko210}, asymmetric channels, 
or the asymmetry introduced by multiple ac drives in a system with 
a symmetric substrate \cite{Hastings111,Hastings12}. 
Ratchets constructed of periodic two-dimensional (2D) arrays of 
asymmetric pinning sites have also been proposed \cite{Zhu13}. 
Recently, both positive and negative vortex rectification have been
experimentally realized in periodic arrays of triangular pinning 
sites \cite{Villegas15}. 
In Ref.~\cite{Villegas15}, the triangles are arranged in a square
lattice with the tips of the triangles oriented in the $+y$ direction.
When an ac force is applied in the $y$ direction at low matching fields,
rectification of the vortex motion in the $+y$ direction occurs over
a range of ac amplitudes, with a peak rectification at a particular
amplitude.  For higher matching fields, there is a $-y$ rectification
at lower ac drives, followed by a $+y$ rectification at higher
ac drives. This change in the rectification direction
is explained in Ref.~\cite{Villegas15} as originating from the separate motion
of interstitial vortices at low drives, giving $-y$ rectification, and
the motion of the pinned vortices at higher drives, producing $+y$
rectification.
It is not clear how the motions of these two vortex species can be fully
separated, especially in the presence of thermal fluctuations, and thus
a clearer picture of the vortex dynamics producing the rectification is
needed.
 
In this work we present simulations of
vortices interacting with 2D square arrays of triangular pinning sites 
for parameters corresponding to the recent experiments 
of Ref.~\cite{Villegas15}. 
We find a $+y$ rectification at low ratios $n$ of the number of
vortices to the number of pinning sites,
with a maximum $+y$ dc flux as a function of
ac amplitude. When interstitial vortices are present,
the initial 
rectification is in the $-y$ direction, and is followed by a 
crossover to $+y$ rectification at higher ac amplitude, 
as seen in experiments. 
We find that for large ac amplitudes, when the motion of the vortex
lattice becomes elastic, the rectification switches to the
$-y$ direction, which is explained in terms of channeling effects. 
We also predict a novel {\it transverse} ratchet effect 
where a dc motion is generated in the direction transverse to the
applied ac drive.
All of the effects we describe here should also occur for 
repulsively interacting colloidal particles interacting with 2D triangular
traps. 

We consider a thin film superconductor containing a square array of 
$N_p=64$ triangular pinning sites  
with periodic boundary conditions in the $x$ and $y$-directions.
We add $N_{v}=nN_p$ vortices to the sample, each of which obeys the
overdamped equation of motion  
\begin{equation}
\eta\frac{d {\bf r}_{i}}{dt} = {\bf f}_{i}^{vv} +
{\bf f}_{p} + {\bf f}_{AC} + {\bf f}_{T} \ .
\end{equation}
Here the
vortex-vortex interaction force, appropriate for a thin film 
superconductor, is ${\bf f}_{i}^{vv} = \sum_{j \neq i}^{N_{v}}
(\Phi_0^2/\mu_0\pi\Lambda){\hat r}/r_{ij}$,
where $\Phi_0$ is the elementary flux quantum and $\Lambda$
is the thin film screening length \cite{Clem16}.
We 
use a fast summation method to evaluate this long-range interaction
\cite{Jensen17}. 
The pinning force ${\bf f}_p$ comes from 
a square array
of equilateral triangles with one vertex pointing in the $+y$ direction,
as shown in Fig.~1.
Every triangular

\begin{figure}
\center{
\epsfxsize=3.5in
\epsfbox{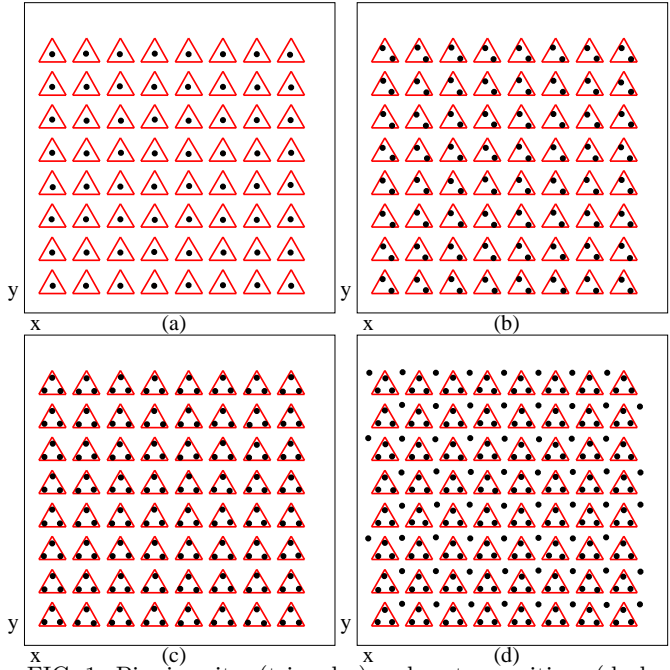}}
\caption{Pinning sites (triangles) and vortex positions (dark circles)
at the matching fields $n=$ (a) 1, (b) 2, (c) 3, and (d) 4.
}
\end{figure}

\noindent
pin is modeled as three half-parabolas of equal
strength, each of which attracts the vortex to a line passing 
through the center of the triangle and parallel to one of the sides.  
The pinning force is cut off at the edge of the triangle.
The applied ac force 
${\bf f}_{AC} = A\sin(\omega t){\bf \hat r}$ acts in the
direction perpendicular to an applied ac current.
Here we consider either $f_{AC}^y=A\sin(\omega t){\bf \hat y}$
or $f_{AC}^x=A\sin(\omega t){\bf \hat x}$, with no mixtures of ac drives.
Temperature is modeled as random thermal kicks with the property
$<{\bf f}_{T}(t)> = 0$ and
$<{\bf f}_{T}(t){\bf f}_{T}(t^{\prime})> = 2\eta k_{B}T\delta(t - t^{\prime})$.

We match our parameters to those of the experiment in Ref.~\cite{Villegas15},
performed in Nb films
with $\eta=1.4\times 10^{-12}$ Ns/m.  We take $T/T_c=0.98$,
giving $f_T=0.46f_0$, where $f_0=\phi_0^2/2\pi\mu_0\lambda^3=1.09\times
10^{-5}$ N/m.  At this temperature, the London penetration depth
$\lambda=\lambda(0)/(1-(T/T_c)^2)^{1/2}=368$ nm, so our pin spacing is
$2.09\lambda$ in the $x$ direction and $2.03\lambda$ in the $y$ direction, and
the pins are $1.68\lambda$ on a side.
In the experiment, each pin held a maximum of
three vortices, so to match this we set the pinning strength to 
$f_p=1.05f_0$. 
We fix $\omega=78$ kHz, higher than experiment due to the limitation of
simulation time; however, since the behavior of the system is controlled by
the ratio $A/\omega$, we compensate by taking $A$ larger than in
experiment.  We note that we find the same generic behaviors
for other parameters,
including a pinning array that is triangular rather than square.

In Fig.~1 we show the vortex positions and pin
locations at $A=0$ for a system with a square
array of triangular pins at fillings $n=1$, 2, 3, and 4.
A global symmetry 
breaking occurs at $n=2$ between the two
possible arrangements of the vortices allowed by the square
pinning 

\begin{figure}
\center{
\epsfxsize=3.5in
\epsfbox{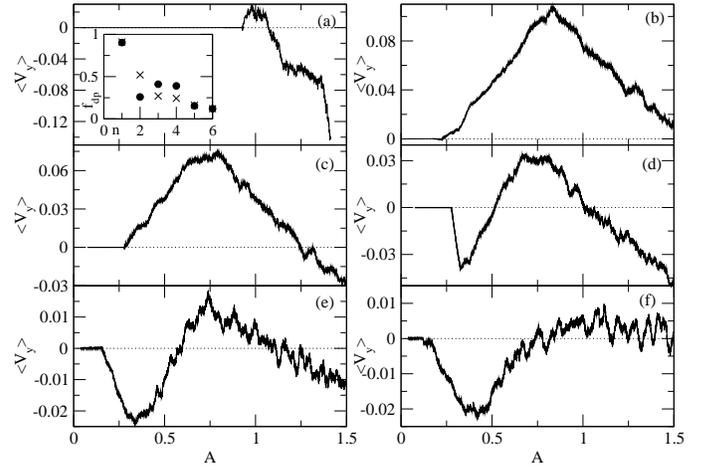}}
\caption{
Net dc velocity $<V_y>$ in units of m/s as a function of applied
ac current $A$ in units of $f_0$
for $n=$ (a) 1, (b) 2, (c) 3, (d) 4, (e) 5, and (f) 6.
Inset to (a): dc depinning force $|f_{dp}|$ for $+y$ (circles)
and $-y$ (x's) directions.
}
\end{figure}

\noindent
lattice.
All of the vortices sit in the pins for $n \le 3$, but
for $n \ge 4$ the 
vortex-vortex interaction is 
strong enough that some vortices move to the 
interstitial regions. 

We first consider the dc depinning force $f_{dp}$ in the positive and negative
$y$ directions by applying a dc driving force of increasing magnitude
and measuring the average vortex velocity $<V_y>$.  
In the inset to Fig.~2(a) we plot $|f_{dp}|$ as a function of 
$n$.  At $n=1$ the depinning is symmetric in the $\pm y$
directions, as expected given our model for the pinning.  
For $n=2$, $|f_{dp}^{-y}|$ is twice as large as $|f_{dp}^{+y}|$,
due to the fact that the vortices align vertically for $+y$ driving, 
and 
horizontally for $-y$ driving.
For $n=3$ and 4, 
$|f_{dp}^{+y}|$ is 25\% larger than
$|f_{dp}^{-y}|$.
Here, the vortex at
the top of each pin acts to assist in the depinning of the vortices
along the bottom of the pin for $-y$ driving. 
For $n>4$, the depinning is dominated by the interstitial vortices and the
asymmetry in the depinning is lost.
The depinning asymmetry at $n=2$ to 4 
should be large enough to be observable experimentally.
Depinning in the $\pm x$ directions is symmetric for all $n$.
Asymmetric depinning at noninteger $n$ was observed in
Ref.~\cite{ZhuV}.

We next consider the effect of an applied ac drive 
$f_{AC}^y$.
We monitor the net dc velocity $<V_y>$
at fixed $\omega$ and sweep the ac amplitude $A$,
averaging $<V_y>$ for 20 periods at each 
increment of $A$.
In Fig.~2(a) we plot $<V_y>$
vs $A$ for $n = 1$.
For $A<0.9$, there is no 
net flow in $x$ or $y$, indicating that the vortices are still pinned.
For $A= 0.9$ a finite $+y$ dc velocity appears
with no net $x$ dc velocity, corresponding to a $+y$
rectification. As $A$ increases 
further, $<V_{y}>$ increases, reaching a maximum at 
$A = 0.99$ and then 
dropping back to zero at $A=1.05$. 
This behavior matches that seen in the experiments of Ref.~\cite{Villegas15}.  
We find no steps in the $+y$ rectification regime, in agreement
with the experiments, 
but in disagreement 
with recent $T=0$
simulations of a very small system for asymmetric pinning sites
that predicted 
steps in the velocity vs 

\begin{figure}
\center{
\epsfxsize=3.5in
\epsfbox{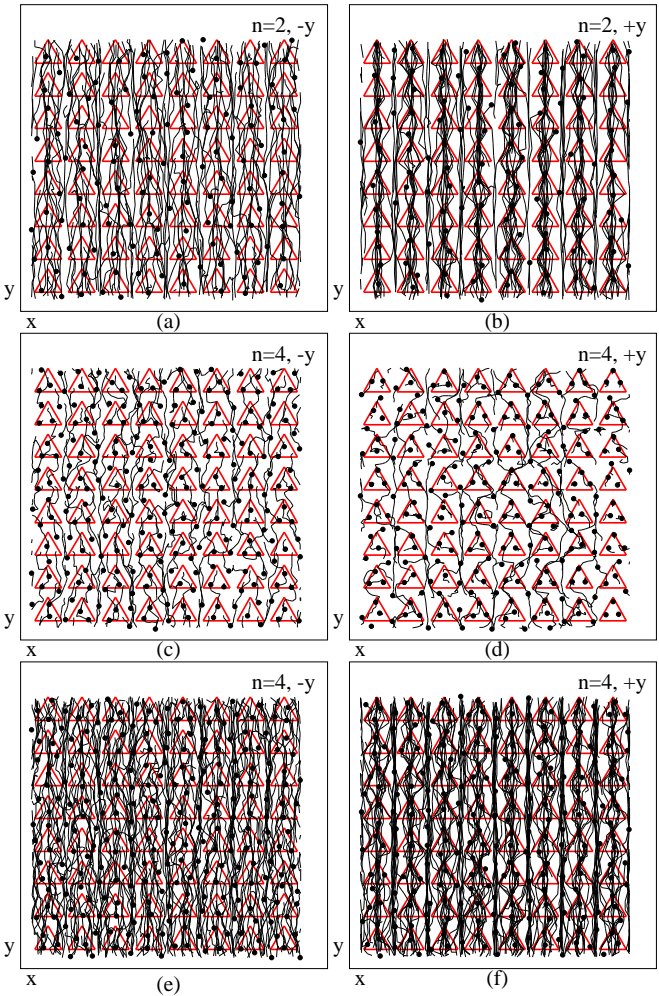}}
\caption{Pins (triangles), vortices (dots), and vortex
trajectories during $1/4$ of the driving period (lines) for: (a) 
$N=2$, $A=1.0$, $-y$ portion of drive cycle, (b) $+y$
portion of same cycle; (c) $N=4$, $A=0.35$, $-y$ portion
of drive cycle, (d) $+y$ portion of same cycle; (e) $N=4$,
$A=0.75$, $-y$ portion of drive cycle, (f) $+y$
portion of same cycle.
}
\end{figure}

\noindent
$f_{AC}$ curve \cite{ZhuE18}.
The transport 
does not occur in an 
organized step like 
flow, but is instead a stochastic process, as we 
will show in more detail in Fig.~3. 
In Fig.~2(b) and Fig.~2(c) we plot $<V_{y}>$ for $n = 2$
and 3, respectively. These curves are similar to the
$n = 1$ case, with an initial pinned phase 
followed by $+y$ rectification that goes through
a maximum 
with increasing $A$. 

In Fig.~2(d) we show $<V_{y}>$ for the case of $n = 4$ when interstitial
vortices are present, as seen in Fig.~1(d). Here there is an initial
{\it negative} rectification regime in the $-y$ direction, 
followed by a crossover to a $+y$ rectification
as seen in experiments \cite{Villegas15}. In Fig.~2(e,f) we plot the cases for 
$n = 5$ and $n = 6$, respectively, which show a similar rectification
to that in Fig.~2(d). 
In all of these cases, a combination of interstitial and pinned
vortices are present.
As $n$ is increased further, the ratcheting effects are gradually reduced
and become indistinguishable from thermal noise. 

For $n=1$ in Fig. 2(a), at higher $A>1.05$, 
$<V_y>$ becomes {\it negative}, indicating $-y$ rectification
at high drives.  This reversal occurs for all fillings but is most
pronounced for $n=1$ and $n=4$, when the onset of the reversal falls
at lower values of $A$.  We predict that
this effect should be easily observable in experiment,
although it was not reported in Ref.~\cite{Villegas15}.
The high-drive $-y$ rectification 
originates when the $-y$ vortex motion 
occurs through interstitial channels
while the $+y$ vortex motion becomes localized on top
of a column of pinning, giving a larger 
drag during the
$+y$ portion of the cycle.

To understand the role of interstitial and pinned vortices in the
rectification process, we examine the dynamics of the vortices. 
In Fig. 3(a,b) we illustrate a typical example
of vortex motion leading to $+y$ rectification at $n=2$ and $A=1.0$.
Fig. 3(a) shows the vortex trajectories 
during 1/4 of the drive cycle when the drive is in the $-y$ direction.
At these temperatures, none of the vortices remain pinned.
The vortex motion is uncoordinated and the vortices wander 
stochastically as they pass through the pinning.  
In contrast, during the $+y$ portion of the
cycle shown in Fig. 3(b), the vortices are channeled both within and
between the columns of pinning sites, and there is little wandering
in the $x$ direction.  This more focused motion leads to a greater
overall displacement in the $+y$ direction, and a net $+y$ rectification.
Similar motion occurs for $n=1$ and $n=3$, 
except that for $n=1$ all of the motion is confined
to the pinning channel and there is no interstitial motion, 
so the net rectification is correspondingly reduced as seen in Fig.~2(a).

At $n=4$ and above, when interstitial vortices
are present, $-y$ rectification occurs at low drives, as shown
in Fig. 2(d-f).  In Fig. 3(c-d) we illustrate an
example of vortex motion in the $-y$ rectification regime 
for $n=4$ and $A=0.35$.
Fig. 3(c) shows that the vortices meander moderately in the $x$ direction
during the $-y$ portion of the drive cycle.
Note that none of the vortices remain pinned; instead, {\it all} of the 
vortices are participating in the motion, with pinned and interstitial 
vortices switching places frequently.  
During the $+y$ portion of the drive cycle shown
in Fig. 3(d), the drive is too low to overcome the pinning force and allow
the formation of channels of the type seen in Fig. 3(b).  
Instead, the $+y$ vortex flow is strongly diverted into the $x$ 
direction by the pins, reducing the net motion in the $+y$ direction,
and resulting in an overall $-y$ rectification.
As $A$ is further increased, the vortices
begin to overcome the pinning in the $+y$ direction and form channels
during the $+y$ portion of the cycle,
as illustrated in Fig. 3(f) for $n=4$ and $A=0.75$.  
This gives $+y$ rectification just as in Fig. 3(a-b).
As $n$ increases above 4, the channels become more closely
packed and the amount of vortex wandering in the $x$ direction during the
$-y$ portion of the cycle decreases, 
progressively lowering the net $+y$ 
rectification.
The onset of the $-y$ and $+y$ rectification phases also drops to 

\begin{figure}
\center{
\epsfxsize=3.5in
\epsfbox{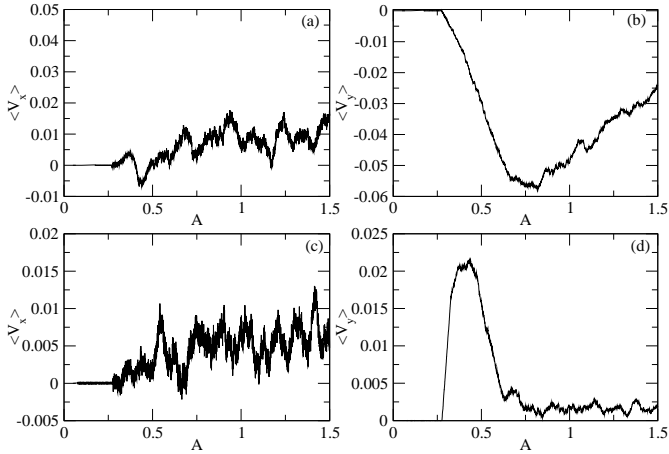}}
\caption{
Transverse rectification for $f_{AC}^x$.
(a) $<V_x>$ for $n=2$; (b) corresponding $<V_y>$ with $-y$
rectification.  (c) $<V_x>$ for $n=4$; (b) corresponding $<V_y>$
with $+y$ rectification.
}
\end{figure}

\noindent
lower $A$ 
with increasing $n$ since the effectiveness of the pinning decreases 
with increasing vortex density.

In Fig. 4 we illustrate $<V_x>$ and $<V_y>$
when an ac drive $f_{AC}^x$ is applied in the $x$ direction
for $n = 2$ and 4. The net $x$ velocity is zero but we find
a {\it transverse} rectification in the $y$ direction: $-y$
for $n=2$ and $+y$ for $n=4$.
The magnitude of the
rectification is comparable to that seen for $f_{AC}^y$, and it
should thus be experimentally observable.
The transverse ratchet effect is
produced by the interaction of the vortices with the pinning sites.
For $n=2$, the vortices channel along the rows of pins and are 
deflected downward by the pinning sites, producing $-y$ 
rectification.  Similar motion occurs for $n=3$.
For $n=4$ to 6, the interstitial vortices dominate the motion through
channels that pass along the tips of the triangles,
although all vortices are being pinned and depinned.  
The interstitial vortices are deflected upward
by the vortices inside the pins, producing $+y$ rectification.
Since the transverse motion does not require breaking of the $x$ direction
symmetry of the system, it can also be observed for a dc $x$ 
direction drive.

In conclusion, we have conducted simulations of
vortices in thin film superconductors interacting with triangular 
pinning sites for parameters relevant to recent experiments. 
We show that both a $+y$ and $-y$ rectification
can occur at depinning 
depending on whether interstitial vortices are present, in agreement with 
experiment.
We find dc depinning anisotropy for $n=2$ to 4, but observe that
even when the dc depinning is {\it not} anisotropic,
rectification can still occur
due to the fact that the vortex dynamics differs under ac and dc drives.   
Rectification in the $+y$ direction occurs when the ac drive
overcomes the pinning strength and vortex channels form that flow
in the $+y$ direction.
When interstitial vortices
are present at zero ac drive, an initial $-y$ rectification occurs
at lower ac drives due to scattering of the $+y$ vortex motion
by the pinning sites.
This occurs for $n = 4$ and above,
in agreement with experiments.
We also predict that a novel transverse ratchet effect should occur 
when the ac drive is applied in the $x$ direction.  Here, the 
rectification is in the $y$ direction.
Our results should be generic to any type of repulsively
interacting particles moving through triangular traps and may have
potential applications for species fractionalization in colloidal
systems. 

This work was supported by the U.S. DoE under Contract No.
W-7405-ENG-36.


\end{document}